\documentclass[aps,twocolumn,english,superscriptaddress,floatfix,longbibliography]{revtex4-2}
\usepackage{graphicx,amsmath,amssymb,natbib,color,soul,easyReview,comment,lineno,appendix}

\usepackage[version=4]{mhchem}
\usepackage[absolute,overlay]{textpos} % Add the textpos package for positioning
\usepackage{graphicx}% Include figure files
\usepackage{dcolumn}% Align table columns on decimal point
\usepackage{bm}% bold math
\usepackage[utf8]{inputenc}
\usepackage[T1]{fontenc}
\usepackage{mathptmx}
\usepackage{etoolbox}
\usepackage{hyperref} %for DOI link 

\usepackage{color,xcolor}
\definecolor{red}{rgb}{0.8, 0.0, 0.0}
\definecolor{darkgreen}{RGB}{0.0, 0.0, 0.0}

\begin{document}

\title{Efficient training of machine learning potentials for metallic glasses: CuZrAl validation}

\author{Antoni Wadowski}
\affiliation{NOMATEN Centre of Excellence, National Center for Nuclear Research, ul. A. So\l{}tana 7, 05-400 Swierk/Otwock, Poland.}
\affiliation{Faculty of Materials Science and Engineering, Warsaw University of Technology, Wo\l{}oska 141, 02-507 Warsaw, Poland.}

\author{Anshul D.S. Parmar}
\email{Anshul.Parmar@ncbj.gov.pl}
\affiliation{NOMATEN Centre of Excellence, National Center for Nuclear Research, ul. A. So\l{}tana 7, 05-400 Swierk/Otwock, Poland.}

\author{Filip Kaśkosz}
\affiliation{NOMATEN Centre of Excellence, National Center for Nuclear Research, ul. A. So\l{}tana 7, 05-400 Swierk/Otwock, Poland.}

\author{Jesper Byggm\"astar}
\affiliation{Department of Physics, P.O. Box 43, FI-00014 University of Helsinki, Finland.}

\author{Jan S. Wr\'obel}
\affiliation{Faculty of Materials Science and Engineering, Warsaw University of Technology, Wo\l{}oska 141, 02-507 Warsaw, Poland.}

\author{Mikko J. Alava}
\affiliation{NOMATEN Centre of Excellence, National Center for Nuclear Research, ul. A. So\l{}tana 7, 05-400 Swierk/Otwock, Poland.}
\affiliation{Aalto University, Department of Applied Physics, PO Box 11000, 00076 Aalto, Espoo, Finland.}

\author{Silvia Bonfanti}
\thanks{Silvia.Bonfanti@unimi.it}
\affiliation{Center for Complexity and Biosystems, Department of Physics "Aldo Pontremoli", University of Milan, Via Celoria 16, 20133 Milano, Italy.}
\affiliation{NOMATEN Centre of Excellence, National Center for Nuclear Research, ul. A. So\l{}tana 7, 05-400 Swierk/Otwock, Poland.}

\date{\today}

\begin{abstract}

Interatomic potentials are key to uncovering microscopic structure–property relationships, essential for multiscale simulations and high-throughput experiments. 
%For metallic glasses, their disordered structure makes potential development especially challenging; as a result, chemistry-specific interaction potentials for this important class of materials are often missing. 
For metallic glasses, their disordered atomic structure makes the development of potentials particularly challenging, resulting in the scarcity of chemistry-specific parametrizations for this important class of materials.
We address this gap by introducing an efficient methodology to design machine learning interatomic potentials (MLIPs), benchmarked on the CuZrAl system. Using a Lennard-Jones surrogate model, swap-Monte Carlo sampling, and single-point Density Functional Theory (DFT) corrections, we capture amorphous structures spanning 14 decades of supercooling. These representative configurations, competing with the experimental time scale, enable robust model training across diverse states, while minimizing the need for extensive DFT datasets. The resulting MLIP matches the experimental data and predictions of the classical embedded atom method (EAM) for structural, dynamical, energetic, and mechanical properties. This approach offers a scalable path to develop accurate MLIPs for complex metallic glasses, including emerging multi-component and high-entropy systems.

%Interatomic potentials play a vital role in revealing microscopic details and structure-property relations, which are fundamental for multiscale simulations and to assist high-throughput experiments. 
%For metallic glasses, developing these potentials is challenging due to the complexity of their unique disordered structure. As a result, chemistry-specific interaction potentials for this important class of materials are often missing.
%Here, we solve this gap by implementing an efficient methodology for designing machine learning interatomic potentials (MLIPs) for metallic glasses, and we benchmark it with the widely studied CuZrAl system. 
%By combining a Lennard-Jones surrogate model with swap-Monte Carlo sampling and Density Functional Theory (DFT) corrections, we capture diverse amorphous structures from 14 decades of supercooling. These distinct structures, competing with the experimental time scale, provide robust and efficient training of the model and applicability to a broad spectrum of supercooled states. This approach reduces the need for extensive DFT optimization datasets while maintaining high accuracy. 
%The MLIP from the present methodology shows results comparable to experimental observations and classical embedded atom methods (EAM), which are available for CuZrAl to predict structural, energetic, and mechanical properties. This work paves the way for the development of new MLIPs efficintly for complex metallic glasses, including emerging multicomponent and high entropy metallic glasses.
\end{abstract}

\maketitle
\section{Introduction}
Metallic glasses (MGs) are an extraordinary class of materials composed of metallic elements arranged in a disordered atomic structure. This unique structure gives them a range of exceptional properties, such as high strength, hardness, and elasticity~\cite{greer1995metallic, KruzicReview2016, suryanarayana2017bulk}. Consequently, MGs are increasingly being used in many different fields, e.g., electronics, biomedical engineering, nanotechnology, and aerospace~\cite{gao2022recent,greer2023metallic,sohrabi2024manufacturing}.
However, the disordered nature of MGs is also a limitation, giving rise to a complex and rugged potential energy landscape (PEL)~\cite{debenedetti2001supercooled,bonfanti2017methods} further complicated by the vast compositional variability~\cite{forrest2023evolutionary}.  
The composition space of metallic glasses remains largely unexplored due to the vast number of possible elemental combinations and the complexity of their mixing behavior.
Predicting their properties and exploring optimal compositions is therefore a challenging task. For this reason, the discovery of novel MGs has traditionally relied on intensive experimental trials and errors~\cite{telford2004case}, only recently supplemented with machine learning methods~\cite{forrest2023evolutionary,merchant2023scaling,makinen2024bayesian}, combined with high-throughput experimentation~\cite{sarker2022discovering,wu2024high}. Improving the microscopic understanding of MGs would significantly advance their exploration. 

To efficiently explore atomic-scale structures, \textit{in silico} calculations have become fundamental, offering microscopic insights that are often inaccessible due to experimental limitations~\cite{binder2011glassy}. Accurate interatomic potentials that mimic atomic interactions and address compositional complexity are central to these simulations. 
However, chemistry-specific potentials for metallic glasses are often lacking, due to their disordered structures and compositional complexity. 
Density functional theory (DFT) and \textit{ab initio} simulations accurately describe atomic interactions for various compositions. However, their computational cost restricts applicability to small systems and short timescales, hindering efficient sampling of the rugged PEL of MGs.  
%To efficiently explore atomic-scale structures, \textit{in silico} calculations have become fundamental, offering microscopic insights that are often inaccessible due to experimental limitations~\cite{binder2011glassy}. 
%In particular, the development of accurate interatomic potentials that mimic atomic interactions is central to these computer simulations. 
%However, chemistry-specific potentials for MGs are often missing, due to their disordered structure and composition complexity.
%The density functional theory (DFT) and \textit{ab initio} simulations accurately describe atomic interactions; however, computational cost restricts applicability to small systems and timescales, hindering the PEL sampling.  
%As an alternative route to precisely describe atomic interactions, density functional theory (DFT) and \textit{ab initio} calculations can be considered.  
%Nevertheless, their computational cost restricts applicability to small systems and timescales, and cannot capture the rugged energy landscape of MGs.
Simplified model potentials, such as Lennard-Jones (LJ) binary mixtures~\cite{kob1995testing} or polydisperse systems~\cite{ninarello2017models}, can satisfactorily describe generic glassy behavior but are not designed to capture specific chemical concentrations or composition-dependent local properties~\cite{voigt2024differences}.
For example, they do not explain why substitutional metallic glasses, where one metal replaces another of similar atomic radius, exhibit different dynamical behaviors in experiments~\cite{voigt2024differences}. 
The semi-empirical, embedded atom method (EAM) based interaction provides a physically accurate and computationally efficient description for MGs. However, accuracy is limited to specific compositions, and the reparameterisation-complexity reduces its transferability to realistic, multicomponent systems~\cite{daw1993embedded, muser2023interatomic}.
%Finally, classical potentials for metallic glasses, such as embedded atom method (EAM), tuned semi-empirically, are computationally efficient even for large bulk systems. However, EAM potentials are still limited in accuracy, available only for specific compositions and present challenges in parameterization that reduce their transferability for the study of realistic, multicomponent systems~\cite{daw1993embedded,muser2023interatomic}.
These limitations led to the development of machine learning interatomic potentials (MLIPs), which allow us to approximate the PEL with near-DFT accuracy, while enabling large-scale simulations~\cite{behler2016perspective, deringer2019machine}, applicable to disordered systems~\cite{bartok2018machine,deringer2017machine,sosso2012neural,sosso2012neural,xie2021neural,ZHAO2023112012,bertani2024accurate}.  
%MLIPs have been successfully employed for a range of disordered systems~\cite{bartok2018machine,deringer2017machine,sosso2012neural,sosso2012neural,xie2021neural,ZHAO2023112012,bertani2024accurate}. 
However, MLIPs also have two major drawbacks: (i) Their accuracy and robustness rely heavily on the quality of the training data, which for glasses is often limited to time scales much shorter than those observed experimentally. (ii) The need for large datasets and high dimensionality increases computational complexity, raising challenges for their transferability and overall robustness~\cite{montes2022training, qi2024robust,wang2024machine}. 

To address these shortcomings, this work proposes an efficient methodology for tailoring MLIPs for metallic glasses. It combines a computationally inexpensive Lennard-Jones surrogate model, accelerated sampling via swap-Monte Carlo, and additional single-point DFT corrections to generate distinct amorphous structures at timescales comparable to experiments. As a result, this approach achieves both physical accuracy and computational efficiency for potential training. 
%First, a LJ-surrogate model 
%parametrized with DFT~\cite{Caro_lattice}, 
%whose parameters are tuned to the target system with DFT~\cite{Caro_lattice},
First, we employ a surrogate Lennard-Jones (LJ) potential, using parameters derived from DFT in Ref.~\cite{Caro_lattice}, to represent the target MG system, providing a simple and effective framework to explore the rugged PEL of MGs. 
To extend the range of disordered configurations and access deeply supercooled states, which are otherwise unattainable with conventional simulation methods, non-local moves using swap-Monte Carlo sampling are performed~\cite{ninarello2017models,parmar2020ultrastable}.
Finally, single-point DFT corrections are applied to the obtained LJ-surrogate structures to refine energies and forces with first-principles, capturing realistic chemistry-specific interactions and generating high-accuracy data for training-testing the MLIP.
%This hybrid approach yields realistic amorphous configurations with single-point DFT corrections, eliminating the need for full DFT optimizations and bypassing the most computationally expensive steps in MLIP development for metallic glasses. The result is a general-transferable framework for modeling complex systems.
This hybrid approach bypasses the most computationally expensive aspects of MLIP development for MGs. Realistic amorphous configurations are generated through accelerated sampling of the LJ-surrogate PEL, with single-point DFT corrections refining the structures and eliminating the need for full DFT optimizations. These steps address the challenges of dataset quality and computational cost, resulting in a general-transferable framework for modeling complex disordered systems.
%This hybrid approach bypasses the most computationally expensive aspects of MLIP development for MGs. Realistic amorphous configurations are generated through an accelerated sampling of the effective LJ-surrogate PEL, and single-point DFT corrections refine these structures, eliminating the need for full DFT optimizations. These steps address the challenges of dataset quality and computational cost, resulting in transferable MLIP performance for modeling complex MGs.. 
\begin{figure}%[Htp]
    \centering
    \includegraphics[width=0.5\textwidth]{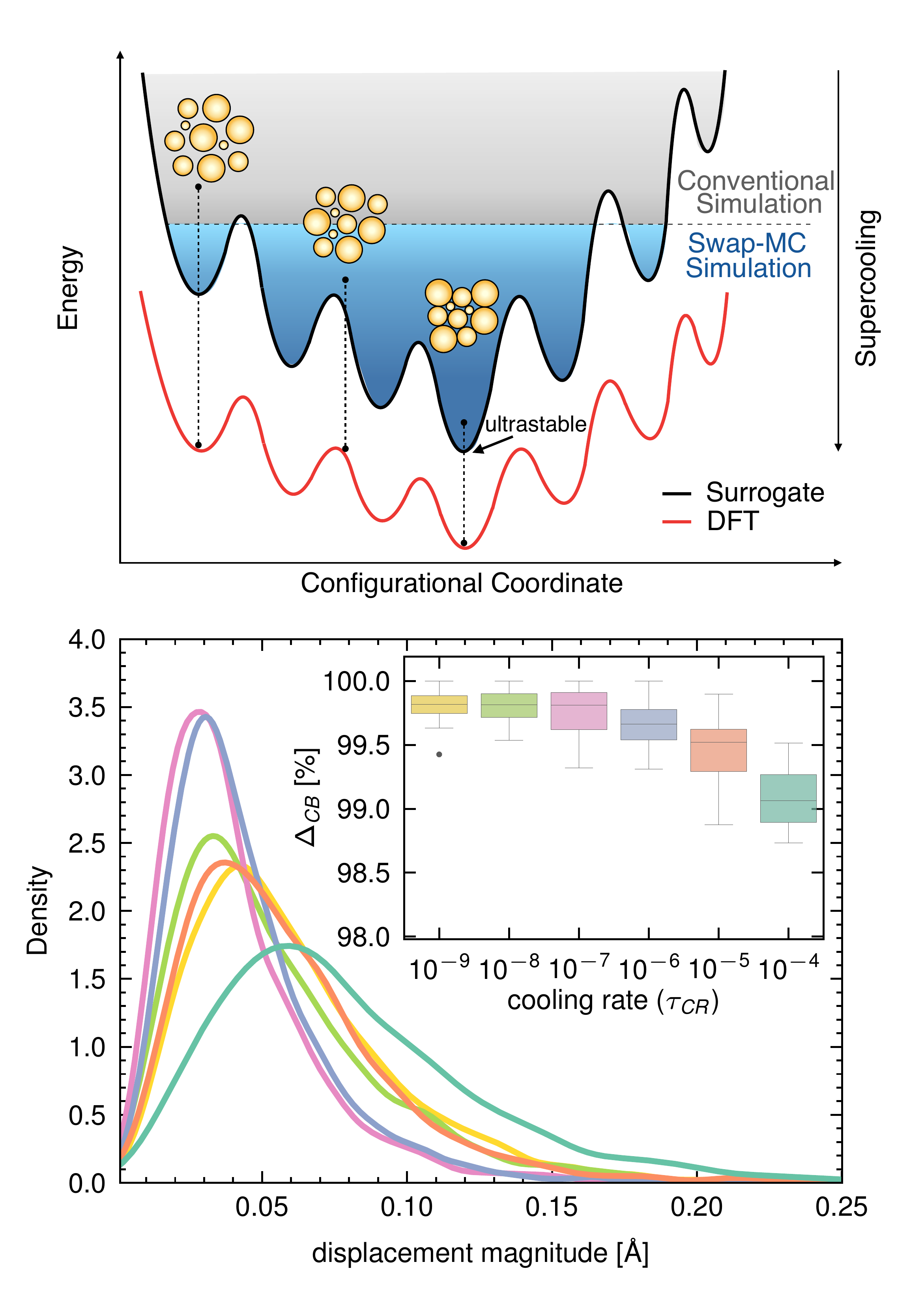}
    %\vspace{-2.mm}
    % \caption{{\bf Schematic of PEL exploration of the LJ-surrogate model with swap-MC, used for accelerated sampling and connection to DFT.} 
    % Swap-MC simulations provide samples from the extended regime of the LJ-surrogate PEL (black line), which are both associated with conventional simulation methods (grey region) and the deeper energy minima from the astronomical time scales (blue region).  
    % Configurations obtained with our LJ-surrogate model (yellow particles as representative) can be directly utilized to investigate the DFT PEL (red line) through single-point DFT corrections. 
    % The methodology bypasses the necessity of expensive DFT optimizations and enables the exploration of the largest range of supercooling.
    %\caption{{\bf PEL exploration of the LJ-surrogate model with swap-MC, used for accelerated sampling and connection to DFT.}
    \caption{\textbf{The LJ-surrogate model enables efficient sampling of deep glassy states and produces configurations that closely match the DFT PEL.} 
    \textbf{a)}~Schematic of swap-MC simulations provide samples from the extended regime of the LJ-surrogate PEL (black line), which are both associated with conventional simulation methods (grey region) and the deeper energy minima from the astronomical time scales (blue region).  
    Configurations obtained with our LJ-surrogate model (yellow particles as representative) can be directly utilized to investigate the DFT PEL (red line) through single-point DFT corrections.
    % The methodology bypasses the necessity of expensive DFT optimizations and enables the exploration of the largest range of supercooling.
    \textcolor{darkgreen}{\textbf{b)}~The particle displacement distributions during the DFT correction of LJ samples, across different cooling rates ($\tau_{CR}$), indicate only minor atomic rearrangements. The inset displays the similarity in the local neighbourhood~$\Delta_{CB}$, which remains above 98\% for all cooling rates. This confirms that the DFT correction does not significantly modify the structure, thereby justifying the transition from the surrogate to the realistic PEL. Each color represents a different LJ cooling rate.}}   
    \label{fig:swap_md}
    \vspace{-0.2cm}
\end{figure} 

To demonstrate the applicability of the proposed methodology, we employ machine learning neuroevolution potentials (NEP)~\cite{Fan_2022, song2024general} to design a new MLIP for the widely studied CuZrAl metallic glass~\cite{inoue2002formation, das2005work, cheung2007thermal, yu2008poisson, pauly2010transformation, poltronieri2023mechanical, alvarez2023simulated, makinen2024avalanches, jiang2024situ}. 
In the following sections, we describe the effectiveness of the LJ-surrogate model and the DFT potential energy landscape (PEL), along with the structural database, the architecture used for MLIP training, and the resulting model performance. Finally, we compare our MLIP against the available EAM potential and experimental data, showing that it successfully reproduces key structural, dynamical, energetic, and mechanical properties.

\section{Efficient DFT database generation via LJ-surrogate model and Swap-MC}
\begin{textblock*}{25cm}(10.85cm,2.1cm) % width of block, x, y position
    \textcolor{black}{\textbf{a)}}
\end{textblock*}
\begin{textblock*}{25cm}(10.85cm,7.8cm) % width of block, x, y position
    \textcolor{black}{\textbf{b)}}
\end{textblock*}
\textit{Energy landscape and database generation---} 
We adopt a general potential energy landscape (PEL) approach to emphasize the structure of our methodology. As shown in Fig.~\ref{fig:swap_md}\textcolor{darkgreen}{a)}, we use an optimized LJ-surrogate model (see Methods) to efficiently explore configurations 
across a wide range of energies, from high-temperature to deeply supercooled states of the PEL. 
Conventional simulations typically sample only shallow energy basins, but by applying non-local swap Monte Carlo moves, we overcome energy barriers and access deeper minima, including ultrastable glassy states. While the LJ parameterization facilitates accelerated sampling, it lacks chemical specificity. Therefore, single-point DFT corrections are applied to refine the surrogate structures, bridging the gap between efficiency and first-principles accuracy.
% This hybrid strategy allows us to efficiently generate realistic and diverse training data for MLIP development.
%\AW{The methodology bypasses the necessity of expensive DFT optimizations and enables the exploration of the largest range of supercooling, critical for generating diverse MLIP training data.}

\textcolor{darkgreen}{Next, we assess the structural similarity between configurations obtained from the LJ-surrogate PEL and those after DFT correction, using two metrics (see Methods): (i) atomic displacements during the DFT correction and (ii) changes in the local environment, as shown in Fig.~\ref{fig:swap_md}b. The displacements remain small relative to the particle diameter, and the fraction of preserved neighbors consistently exceeds 98\% across various cooling rates. \\
These results indicate that the DFT correction introduces only minor adjustments while preserving the overall structural framework. Together, they validate that the surrogate model produces configurations that are physically consistent with those from DFT, supporting the use of the LJ-surrogate PEL for further analysis.}

\textit{Effectiveness of LJ-surrogate model---} First, to overcome the computationally demanding glass-structure generation with DFT calculations, we perform swap-Monte Carlo for the CuZrAl system, interacting via classical Lennard-Jones (LJ) potential serving as a surrogate model. With the LJ parameterization of various components~\cite{Caro_lattice}, Al-atoms facilitate efficient swapping between Cu and Zr particles, which would otherwise be unattainable (see Methods,~\cite{parmar2020ultrastable}). To access distinct parts of the energy landscape, we cool the samples from high ($T^*_{H}=10.01$) to lower temperatures ($T^*_L=0.01$), while linear cooling with $10^4$ to $10^9$ swap-Monte Carlo steps. 

\begin{textblock*}{5cm}(10.8cm,1.9cm) % width of block, x, y position
    \textcolor{black}{\textbf{a)}}
\end{textblock*}
\begin{textblock*}{5cm}(10.8cm,5.3cm) % width of block, x, y position
    \textcolor{black}{\textbf{b)}}
\end{textblock*}

\begin{figure}
    \raggedleft
    \includegraphics[width=0.48\textwidth]{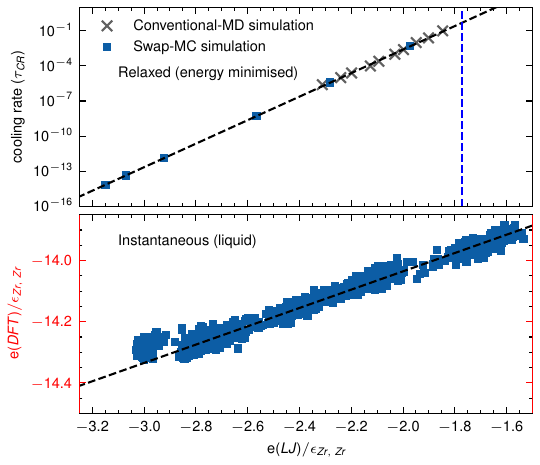}
    \caption{{\bf Times scales of supercooling with swap-MC and LJ-surrogate model, and pathway to the DFT.} \textbf{a)} The energy of the relaxed/minimized structure with MD follows a logarithmic relationship with the cooling rates. The timescales for the swap-MC are marked with the extrapolated energy-logarithmic behavior. The vertical blue line marks the onset of the supercooled regime.
    \textbf{b)} Linear correlation for instantaneous energy of amorphous samples from LJ and DFT showing the relevance of the \textit{surrogate} structural signatures. The black dashed line represents the linear fit to highlight the likeness.} 
    \label{fig:swap_vs_md}
        \vspace{-0.2cm}
\end{figure}

To explore the associated time scales for the accessed energies, we perform standard molecular dynamics (MD) simulation for the same surrogate potential, cooled from $T^*_H$ to $T^*_L$ in MD-time (reduced units) ranging from $10^2$ to $4 \times 10^6$. Figure~\ref{fig:swap_vs_md}a) shows that the energy of the relaxed (inherent) structures follows a logarithmic dependence over the cooling rates for the MD calculations: 
$e_{IS}(LJ) = e_{IS,on}(LJ) + A \log(\tau_{CR}/\tau_{CR,on})$,
where $A$ is the material-specific parameter determining the proclivity for ageing. The $e_{IS,on}$ and $\tau_{CR,on}$ are the reference energy of relaxed structure and cooling rate for the onset of supercooling~\cite{sciortino2002thermodynamics,amir2012relaxations,zhang2022shear}. 
With this empirical observation, we estimate the effective time scale for the relaxed structures (ISs) from the swap-MC. The plot shows a comparative span of energies achieved with the swap-MC and MD calculation, emphasizing that the LJ-surrogate model provides unprecedented access to configurations ranging from high-energy liquid-like states to ultrastable glassy states, spanning over timescales of 14-decades, which is otherwise unfeasible to achieve with conventional MD simulations. 

\textit{DFT corrections---} The output liquid-structures from the LJ-surrogate model and swap-MC simulations are then calculated using single-point DFT. This correction step refines the energy and force accuracy for the structures.
Figure~\ref{fig:swap_vs_md}b) shows the correlation between the instantaneous energy of liquids calculated using the LJ-surrogate model ($e(LJ)$), and the corresponding energy obtained with DFT ($e(DFT)$), for all configurations. The data points show a clear linear relationship, suggesting that the LJ-surrogate model effectively approximates the energy landscape for the sampled liquid configurations and the DFT-relaxed structures (see SI, Fig.~S1), capturing the essential trends of the realistic PEL.  Furthermore, the presence of crystallized samples for the lowest cooling rates highlights extended-sampling, and there is no need to extend the cooling rate further.

\begin{figure*}[t]
    \centering
    \includegraphics[width=0.99\textwidth]{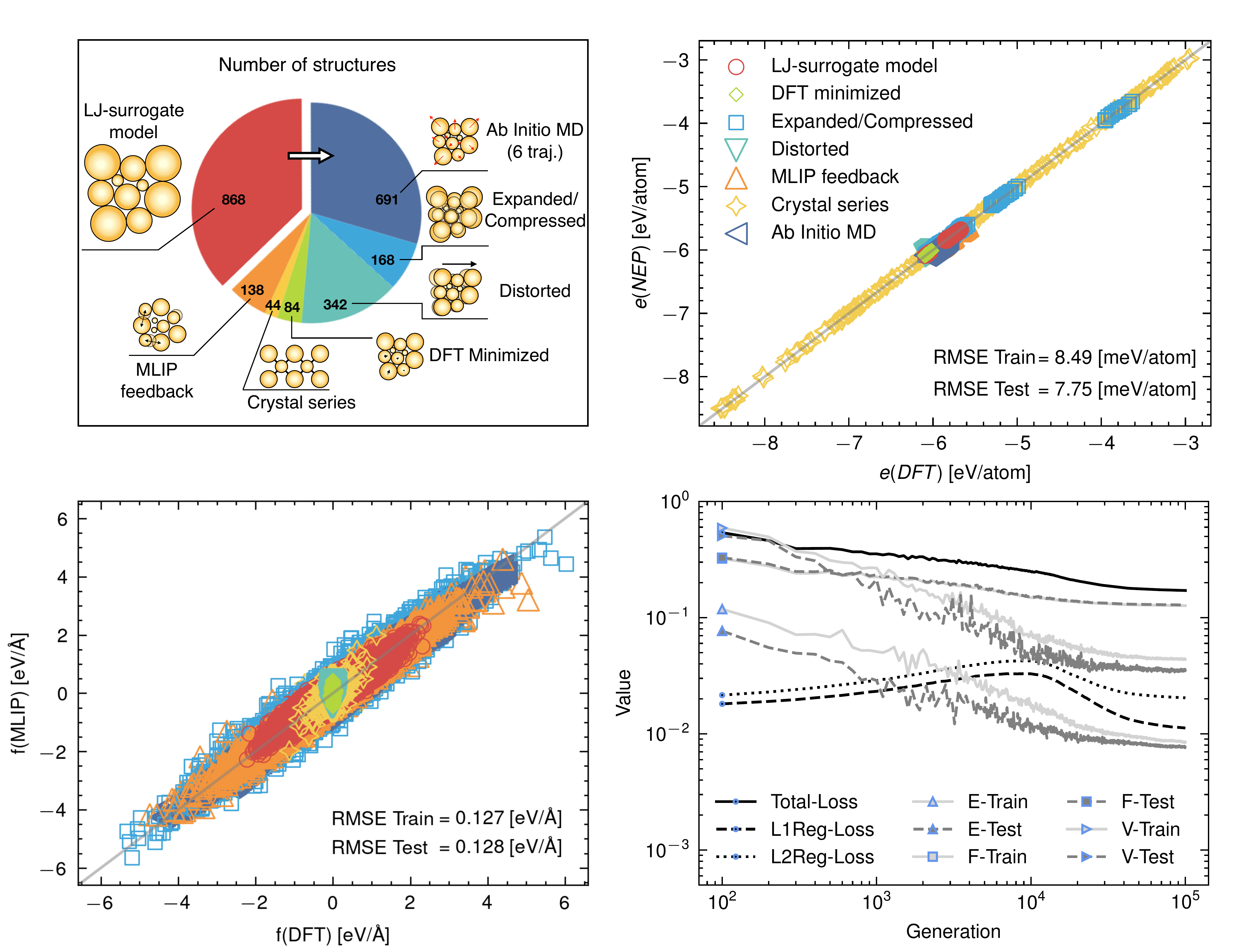}
\caption{\textbf{Minimal DFT database composition to train the MLIP, energy and force validation, and loss convergence.}~~\textbf{a)} Distribution of DFT structures in different subsets within the train dataset.
The largest group consists of the configurations obtained with the LJ-surrogate model, from which the other datasets are derived (as indicated by the arrow). The same color code for structure datasets is used in~\textbf{b)} for MLIP-predicted vs DFT energies, showing very good agreement for both train and test datasets combined in the plot. Similarly, ~\textbf{c)}, shows validation of the force component prediction for the $x$,$y$,$z$ directions combined .~\textbf{d)} Evolution of the MLIP loss, with darker lines for training and lighter for test data. The following total loss function contributions are shown: energy (E) [eV/atom], force (F) [eV/\AA], virial (V) [eV/atom] Root Mean Square Errors, and regularization terms of the parameter vector (L1Reg-Loss, L2Reg-Loss).} 
\label{fig:dft_database_architecture_model_training}
    \vspace{-0.3cm}
\end{figure*}

\textit{DFT database overview---} 
The structures from the LJ-surrogate model are the building blocks of the DFT database development process, the outcome of which is visualized in Fig.~\ref{fig:dft_database_architecture_model_training}{a)}. Samples from the following steps are included to train the MLIP.
(i)~``LJ-surrogate model'' structures, refined with single-point DFT calculations, are used to efficiently sample the PEL of MGs.
(ii)~``DFT minimized'' structures: 10\% of the above samples are further relaxed with DFT to access deeper energy minima, attaining first-principles accuracy. Subgroups of these relaxed structures serve as starting points for steps (iii-v), and (vii), where the model captures key mechanical properties and thermal behavior.
(iii)~``Expanded/Compressed'' samples undergo volumetric changes by iteratively increasing and decreasing all the lattice vectors by 1, 5, and 10\%.
(iv)~``Distorted'' samples are generated by applying strains of $\pm$0.4\% and $\pm$0.8\% in the directions corresponding to the most important~\cite{wrobel2021elastic} components of the stiffness matrix $C_{ij}$: $(ij)$=$[11, 12, 13, 22, 23, 33, 44, 55, 66]$.
(v)~``Ab Initio MD'' structures are heated from 0~K to 2000~K in the NPT ensemble with external pressure of 0~bar.
(vi)~``Crystal series'': Additionally, 44 crystal structures from the Materials Project~\cite{jain_commentary} are included to represent known crystal phases within the CuZrAl system. To improve the MLIP performance for the crystalline phase, each crystal sample is subjected to volumetric changes, as done for the ``Expanded/Compressed'' structures. Consequently, the crystal samples are also distorted by applying strains of ±0.8\%. 
(vii)~``MLIP feedback'': After an initial training of the MLIP, we carry out an active learning-inspired process. We use the trained MLIP to perform equilibration runs, followed by quenching with cooling rates of 10, 100, and 1000~K/ns. The final structures are then computed with single-point DFT and added to the training database. The MLIP is then re-trained to improve the accuracy and robustness of MG modeling.
The MLIP performance is validated against a set of MG structures. To this end, each subset of structures from steps (i-iv)~and (vii)~is randomly divided into train and test datasets, following an 80\% and 20\% split, respectively. Two new \textit{ab-initio} MD trajectories, developed as in step (v), are added to the test dataset. For details of the DFT calculations, refer to the Methods section.
\begin{figure}[htp]
    \centering
\includegraphics[width=0.48\textwidth]{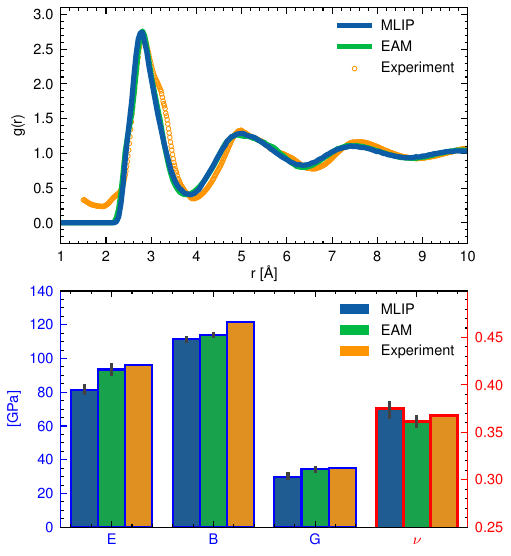}
    % \includegraphics[width=0.43\textwidth]{rdf_mlip_eam_experiment.pdf}
    % \includegraphics[width=0.47\textwidth]{elastic_properties_barlpot.pdf}
    %\caption{\textbf{Comparison between the developed MLIP,  EAM potential~\cite{Cheng_2009} and experimental~\cite{Jiang_Zr} data for the Cu$_{0.46}$Zr$_{0.46}$Al$_{0.08}$ MG.
    \caption{\textbf{{
    Validation of the structural and mechanical properties predicted by the MLIP.}} \textbf{a)} Radial distribution function comparison shows that the MLIP captures coordination shells and peak positions, in agreement with experimental~\cite{Jiang_Zr} and EAM results.
    \textbf{b)} Elastic properties values: Young's ($E$), bulk ($B$), shear modulus ($G$), and Poisson's ratio ($\nu$), are in good agreement with both experimental and \textit{in silico} measurements.}
\label{fig:mlip_application_plot_group} 
\end{figure}

%\begin{textblock*}{1cm}(10.8cm,2cm) % width of block, x, y position
%    \textbf{a)}
%\end{textblock*}
%\begin{textblock*}{1cm}(10.8cm,6.6cm) % width of block, x, y position
%    \textbf{b)}
%\end{textblock*}

\textit{Model training and performance---}
The comparison between DFT energy $e(DFT)$ and predicted energies with trained MLIP $e(MLIP)$ is presented in Fig.~\ref{fig:dft_database_architecture_model_training}b), covering both test and train datasets.
The MLIP demonstrates excellent agreement with DFT, accurately predicting both energies and the various components of forces (see Fig.~\ref{fig:dft_database_architecture_model_training}c) across all datasets, including testing and training.
Figure~\ref{fig:dft_database_architecture_model_training}d) shows the evolution of the total loss function during the MLIP training with NEP, together with the contributions from energy, forces, virials (see Methods and SI, Fig.~S2), and regularization terms of the parameter vector. 
The convergence of the loss function, together with the agreement between training and test sets, indicates that the MLIP generalizes well (see Methods and SI for training parameter). The error in the training and test datasets remains of identical magnitude, suggesting that the MLIP is converged without overfitting, and is ready to be tested further for the physical properties for the experimental and model system. 

\begin{textblock*}{1cm}(2cm,2cm)
    \textbf{a)}
\end{textblock*}
\begin{textblock*}{1cm}(10.6cm,2cm)
    \textbf{b)}
\end{textblock*}
\begin{textblock*}{1cm}(2cm,8.6cm)
    \textbf{c)}
\end{textblock*}
\begin{textblock*}{1cm}(10.6cm,8.6cm)
    \textbf{d)}
\end{textblock*}

\section{Case study - C\lowercase{u}Z\lowercase{r}A\lowercase{l} metallic glass}
The methodology presented in this work is used to develop a MLIP for the widely studied MG composition Cu$_{0.46}$Zr$_{0.46}$Al$_{0.08}$, 
enabling direct comparison with the available EAM potential from Ref.~\cite{Cheng_2009}. In the following, we present a comparative study of the MLIP with DFT, EAM, and experimental results, focusing on structural features, dynamical quantities, mechanical properties, and energies.

\textit{\textcolor{black}{Structure---}}
Firstly, we compare the radial distribution function (RDF), 
($g(r)$), for the MLIP with the existing EAM potential and experimental data. For the MLIP and EAM models, we simulate a system of 1500 particles with 80 independent samples cooled from 2000 K to 300 K in the NPT ensemble at a cooling rate of 100 K/ns. The experimental data are taken from Ref.~\cite{Jiang_Zr} for the cast sample. It is important to note that the cooling rates and sample preparation protocols differ between the \textit{in silico} and experimental settings.
Fig.~\ref{fig:mlip_application_plot_group}a) shows a qualitative comparison between \textit{in silico} and experimental measurements. To quantify the proximity of the simulated RDF to the experimental data, we compute the mean absolute error (or Wasserstein distance) in the range of $r$ [\AA]$\in [2.3,~10]$, obtaining values of 0.091 for MLIP and 0.084 for EAM. 
The close agreement in peak positions and magnitudes indicates that the local structure is well captured. 
The slight shoulder observed in the experimental data may be attributed to the significantly lower cooling rates used during sample preparation, which remain challenging to replicate in simulations.

\textit{Elastic properties---} 
The elastic properties of Cu$_{0.46}$Zr$_{0.46}$Al$_{0.08}$ MG are determined by applying finite structural deformations at 0~K. The resulting stress variations are used to compute the stiffness matrix components, $C_{ij}$. Furthermore, the Young’s modulus ($E$), bulk modulus ($B$), shear modulus ($G$), and Poisson’s ratio ($\nu$) are calculated using the Voigt-Reuss-Hill averaging method~\cite{hill1952elastic, singh2021mechElastic}. The deformation samples are selected from the DFT-minimized subset, specifically filtering structures generated at relatively low LJ cooling rates ($\tau_{CR} < 10^{-4}$). The results, averaged over 85 samples, are presented in Fig.~\ref{fig:mlip_application_plot_group}b) and compared with experimental data. The MLIP demonstrates good predictive accuracy and consistency in the elastic properties of MGs compared to both the EAM potential and experimental data.

%\begin{textblock*}{1cm}(10.8cm,2cm) % width of block, x, y position
%    \textbf{a)}
%\end{textblock*}
%\begin{textblock*}{1cm}(10.8cm,5cm) % width of block, x, y position
%    \textbf{b)}
%\end{textblock*}

\begin{textblock*}{5cm}(1.45cm,1.9cm) % width of block, x, y position
    \textcolor{black}{\textbf{a)}}
\end{textblock*}
\begin{textblock*}{5cm}(1.45cm,6.7cm) % width of block, x, y position
    \textcolor{black}{\textbf{b)}}
\end{textblock*}

{\textit{\textcolor{darkgreen} {Viscosity and specific heat---}} 
{\textcolor{darkgreen} {To evaluate the dynamical behavior of the liquid, we calculate the shear viscosity ($\eta$) using the Green-Kubo relation (see Methods) and compare the results with EAM simulations and existing experimental data~\cite{ding2024mechanical}. 
Figure~\ref{fig:viscosity}(a) shows that the values of the viscosity, obtained using the MLIP potential %are closer to the experiment, suggesting that that MLIP reflects the dynamics more realistically than EAM.
closely match the experimental data throughout the entire temperature range studied (1400–1650~K), in contrast to the EAM potential, which systematically underestimates the viscosity. This suggests that MLIP more effectively captures the temperature-dependent atomic dynamics of the ZrCuAl system, however, further systematic study would be desired.
Turning to the thermodynamic properties, Fig.~\ref{fig:viscosity}(b) presents the specific heat ($C_p$) over a range of temperatures, comparing the MLIP and EAM potentials. From this analysis, the MLIP demonstrates good agreement with EAM, exhibits a physically consistent thermodynamic behavior, and provides a more accurate prediction of the viscosity, supporting the validity of the present model.
%Although the MLIP results exhibit greater uncertainty, this may indicate enhanced sensitivity to local structural variations—an expected feature of machine learning-based potentials. On the other hand, while EAM predictions appear more consistent, they fail to replicate the experimentally observed decrease in viscosity with increasing temperature. Overall, these results underscore the advantage of MLIP in modeling the complex dynamic behavior of metallic glasses, especially in scenarios where traditional empirical potentials are insufficient.
}}

\begin{figure}[ht] 
    \includegraphics[width=0.49\textwidth]{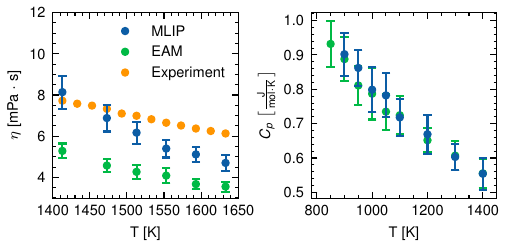}
    \caption{\textcolor{darkgreen}{\textbf{
    Validation of shear viscosity and specific heat values predicted by the MLIP.}~{\bf a)} The MLIP predictions show better agreement with the experimental trend across the temperature range~\cite{ding2024mechanical}, while the EAM systematically underestimates viscosity. 
    {\bf b)} In the thermodynamic analysis, the specific heat shows excellent agreement between MLIP and EAM, indicating physically consistent enthalpy fluctuations. The error bars show the estimate of standard deviation.}} 
    \label{fig:viscosity} 
        % \vspace{-0.1cm}
\end{figure}

\begin{textblock*}{1cm}(10.8cm,11.7cm)
    \textbf{a)}
\end{textblock*}
\begin{textblock*}{1cm}(15.2cm,11.7cm)
    \textbf{b)}
\end{textblock*}

\textit{Shear and energetics---}
For a range of supercooling conditions, we conduct a comparative study between the EAM and MLIP models. The DFT-minimized structures span a broad energy landscape, derived from the LJ-surrogate model (see Fig.~\ref{fig:swap_vs_md}). We perform athermal quasi-static shear simulations~\cite{maloney2006amorphous, bonfanti2019elementary}, where the shear modulus $G$ is determined from the slope of the response curve in the elastic regime, within the strain range $\in [0.004, 0.006]$. The calculated modulus ($G$) values are presented in Fig.~\ref{fig:mlip_eam}a). Notably, the simulation results for MLIP and EAM~\cite{Cheng_2009} are consistent, demonstrating the stability of both potentials across the investigated energy landscape.
\begin{figure}[!htp] 
    \includegraphics[width=0.48\textwidth]{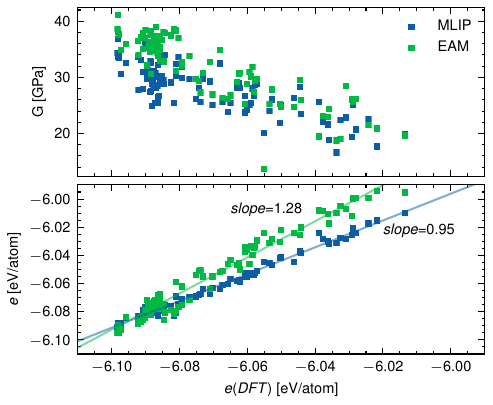}
    \caption{\textbf{{
    The predicted MLIP PEL aligns more closely with DFT than EAM for different supercooling.}}
    \textbf{a)} Shear modulus~$G$ and \textbf{b)} potential energy from MLIP and EAM vs the energy of a DFT-minimized structure. The shear modulus~$G$ systematically increases with the degree of supercooling. Compared with EAM, the MLIP shows a relatively better description of the potential energy with the DFT, i.e., $e(MLIP) \simeq e(DFT)$. 
    Note that to show the energy data within a single plot, the EAM dataset is shifted downwards by~1.11~eV.}
    \label{fig:mlip_eam} 
        \vspace{-0.3cm}
\end{figure}
Lastly, as part of PEL quantification, we compare the minimized structure energies obtained from DFT, MLIP, and EAM models. As shown in Fig.~\ref{fig:mlip_eam}b), the MLIP provides a more consistent description of the energy landscape, closely matching DFT calculations, i.e., $e(\text{MLIP}) \sim e(\text{DFT})$. In contrast, the EAM model systematically overestimates both the energy scale and slope, as highlighted by the linear fit. 
%While MLIP calculations are slower than EAM, they remain significantly faster than \textit{ab initio} methods while maintaining comparable accuracy (see SI, Section~2).
This result underscores the natural advantage of the present methodology, reflecting the fact that the MLIP was directly trained on DFT data, thereby offering a more physically accurate representation of the DFT energy landscape.

\section{Conclusions}
\textcolor{darkgreen}{This work introduces an efficient approach for developing machine learning interatomic potentials (MLIPs) for metallic glasses, applicable to any multicomponent system. As a test case, we apply it to the ternary Cu-Zr-Al system. The method combines a Lennard-Jones surrogate model for accelerated swap Monte Carlo sampling of the potential energy landscape (PEL) with single-point DFT corrections. Structural analysis reveals that these DFT corrections do not significantly alter the underlying LJ-based PEL, reinforcing the conceptual continuity between the surrogate and the true DFT-based landscapes. \\
The training dataset incorporates a wide range of structural features, with particular emphasis on amorphous configurations sampled across 14 decades of supercooling. Conventional sample generation methods that rely on full DFT optimizations often fail to capture the broad spectrum of supercooled states relevant to experimental glasses. In contrast, our method efficiently samples a wider configurational space, learns complex structure–energy relationships, and substantially reduces computational cost while maintaining high accuracy. Furthermore, our database is significantly smaller than many previous multi-element MLIP datasets~\cite{ZHAO2023112012}.\\
As validation and applicability, we demonstrate that the developed MLIP successfully predicts structural, dynamical, thermodynamic, and mechanical properties for the well-known CuZrAl system. }
The MLIP exhibits excellent agreement with experimental data and classical potentials, such as the EAM, effectively capturing the physics of the supercooled CuZrAl system. Although our development and test runs were conducted at a fixed composition, we expect the potential to perform competitively with EAM for other compositions, provided they are not too far from the equiatomic CuZr–Al mixture.\\
The methodology proposed here offers an efficient and transferable framework for the development of MLIPs for more complex MG systems, including multicomponent and emerging high entropy metallic glasses~\cite{afonin2024high}, by using surrogate models, swap-MC techniques, machine learning, and first-principles calculations. It also provides a valuable tool to accelerate the discovery and optimization of new materials with unique structural and mechanical properties.

%This work presents a new efficient approach for developing MLIPs for MGs {\color {darkgreen} for any multicoment system}, as test case applied on ternary CuZrAl system. To this end, we use a combination of a Lennard-Jones surrogate model for accelerated swap-Monte Carlo sampling of PEL, and corrections with DFT. The careful structural evaluation suggests that the single-point DFT correction does not change the initial LJ-sugorate PEL, strengthening the conceptual picture of the transition from LJ to meaningful DFT-based PEL. A variety of structural features were encoded in the initial training structures, with a particular emphasis on amorphous configurations sampled over 14 decades of supercooling. 

\begin{textblock*}{5cm}(1.45cm,1.9cm) % width of block, x, y position
    \textcolor{black}{\textbf{a)}}
\end{textblock*}
\begin{textblock*}{5cm}(1.45cm,5.3cm) % width of block, x, y position
    \textcolor{black}{\textbf{b)}}
\end{textblock*}

%We demonstrate that the developed MLIP successfully predicts structural, dynamical, thermodynamic, and mechanical properties, showing very good agreement with experimental data and classical potentials, such as the EAM. We also note the dependence of the elastic properties on cooling, an important feature that a MLIP should capture.Although our development and test runs were conducted at a fixed composition, we expect the potential to perform competitively with EAM for other compositions, provided they are not too far from the equiatomic CuZr–Al mixture.
%This methodology opens the efficient approach to the development of MLIPs for more complex systems, including multicomponent and emerging high entropy metallic glasses~\cite{afonin2024high}, by efficiently using surrogate models, swap-MC techniques, machine learning, and first-principles calculations. It also provides a valuable tool to accelerate the discovery and optimization of new materials with unique structural and mechanical properties.

\section{Methods}
\textit{Lennard-Jones surrogate model---} 
To develop the interaction potential between elements of multicomponent alloys of Cu$_{0.46}$Zr$_{0.46}$Al$_{0.08}$ consisting of $N=150$ atoms with unit mass (m), we use a surrogate interaction described by the Lennard-Jones (LJ) potential as
\begin{equation}
    e_{\alpha_i,\beta_j} (r_{ij}) = 4\epsilon_{\alpha_i,\beta_j} \left[ \left(\frac{\sigma_{\alpha_i,\beta_j}}{r_{ij}}\right)^{12} - \left(\frac{\sigma_{\alpha_i,\beta_j}}{r_{ij}}\right)^{6} \right], 
\end{equation}
where $\epsilon$ and $\sigma$ are the energy scale and interaction
range, respectively. 
The potential is truncated and shifted at the cutoff distance
$r_{cut,~{ij}} = 2\sigma_{\alpha_i,\beta_j}$.
We specify the atom index by Roman indices and the type by Greek indices. We use interaction diameter as 
$\sigma_{\text{ Zr, Zr}}$=2.932~\AA, $\sigma_{\text{Cu, Cu}}$=2.338~\AA~ and
$\sigma_{\text{Al, Al}}$=2.620~\AA; also energies as $\epsilon_{\text{Zr, Zr}}$=0.409~eV, 
$\epsilon_{\text{Cu, Cu}}$=~0.739~eV and 
$\epsilon_{\text{Al, Al}}$=0.392~eV, respectively. These LJ-equivalent interaction parameters are estimated from the corresponding crystalline structures~\cite{Caro_lattice}. 
Energy (temperatures) and length are in units of  $\epsilon_{\text{Zr, Zr}}$ and $\sigma_{\text{Zr, Zr}}$, respectively. Simulations are performed in the NVT ensemble with number density $\rho^*=1.75$, identical to the mass density from studies~\cite{alvarez2023simulated}. The cross-interaction is modeled with the Lorentz-Berthelot mixing
rules~\cite{Berthelot_comptes}: $\sigma_{\alpha \beta} = (\sigma_{\alpha} +\sigma_{\beta})/2$ and $\epsilon_{\alpha \beta} = \sqrt{\epsilon_{\alpha}\epsilon_{\beta}}$.

\textcolor{darkgreen}{
\textit{Particle's local neighborhood changes---} To quantify changes in each particle's local neighbourhood, we estimate the fraction of neighbour changes per particle following the correction from the LJ-surrogate states to the DFT-minimized states. The bond (nearest-neighbour) network is first determined for the initial LJ-surrogate configuration and then compared to the final DFT-minimized structure.  
We define the bond connectivity for a given particle \( i \) and its neighboring particles \( j \) as those within a distance of \( r_{ij} \leq 4 \)~\AA, corresponding to the first minimum of the pair correlation function \( g(r_{ij}) \).  
\\
The relative structural change during the DFT correction is defined as
\begin{equation}
    {\Delta}_{\mathrm{CB}}^{(i)}=\frac{{n}_{i}(\text{DFT}| \text{LJ})}{{n}_{i}(\text{LJ})},
\end{equation}
where ${n}_{i}(\text{DFT}| \text{LJ})$ represents the number of particle neighbors (i.e., bonds) of particle i in the initial LJ-sample that remain as neighbors after the DFT minimization. Additionally, ${n}_{i}(\text{LJ})$ represents the bond count for the initial LJ sample. Finally, the overall degree of `similarity' in the samples can be defined as 
\begin{equation}
    {\Delta_{CB}} = \left <  \frac{1}{N}\sum_{i=1,N} \Delta^{(i)}_{CB}  \right >,
    \label{eq:particle_neighbor_change}
\end{equation}
where the angular bracket `$\langle \rangle$' represents the averaging over samples for the each cooling rate.
}

\textit{Sampling PEL with swap-Monte Carlo---} 
To explore a wide range of supercooling conditions, we perform Monte Carlo simulations incorporating both particle displacements and exchanges, i.e., swap moves~\cite{ninarello2017models,parmar2020ultrastable}. A single Monte Carlo step consists of \( N \) moves, with 80\% translation and the remaining being swap moves; timescales are reported in this unit. For translation moves, a particle is randomly selected and displaced by a vector chosen within a cube of size \( \delta r_{\max} = 0.15 \).  
For non-local moves, Cu-Zr swaps are effectively rejected due to the significant size mismatch. However, Al atoms, with their intermediate diameter, provide a viable pathway for efficient swap moves~\cite{parmar2020ultrastable}. A randomly selected Al atom is swapped with either a Zr or Cu atom, following the sequence Zr \( \leftrightarrow \) Al \( \leftrightarrow \) Cu.  
Both types of Monte Carlo moves are accepted based on the {Metropolis acceptance rule}, ensuring detailed balance.  To access various regions of the energy landscape, the system is cooled from a high temperature \( T^*_{H} = 10.01 \) to a low temperature \( T^*_{L} = 0.01 \), with cooling rates ranging from $10^4$ to  $10^9$  Monte Carlo moves.

\textit{Estimating the supercooling---} To quantify the degree of supercooling, and associated time scales with the swap-Monte Carlo, we perform conventional molecular dynamics simulations. Similar to Monte Carlo protocols, the samples are cooled form $T^*_{H}=10.01$ to $T^*_{L}=0.01$ with MD time $t^*$(= $\sigma_{\text{Zr, Zr}}\sqrt{(m_{\text{Zr}}/\epsilon_{\text{Zr, Zr}})}$, in reduced units) ranging from $10^2$ to $4\cdot10^4$. The time scales for supercooled swap-Monte Carlo samples are identified with the "logarithmic" energy profile against cooling with molecular dynamics~\cite{sciortino2002thermodynamics}.
We estimate the onset of the supercooled dynamics by looking at the deviation from the Arrhenius behavior at the high-temperature equilibrium dynamics~\cite{kivelson1995thermodynamic}. Which defines the onset temperature ($T_{on}=2.09$), and the corresponding energy minimum $e_{IS, on}(LJ)$ marks the onset of the supercooled regime.  

\textit{DFT Calculations---} 
Each DFT computation included 150 atoms, meeting the requirement of minimum supercell size for MGs~\cite{Holmstrom_structure}. 
Vienna Ab initio Simulation Package (VASP) version 6.3.2 \cite{kresse1996efficiency, kresse1996efficient} was used to perform DFT calculations. The functional used was the projector augmented wave (PAW) Perdew–Burke–Ernzerhof (PBE) \cite{blochl_projector, kresse_from, perdew_generalized}. The cutoff energy was equal to 450 eV. The Monkhorst–Pack mesh \cite{monkhorst1976special} of k points in the Brillouin zone was used, with a k-mesh spacing of 0.162 Å$^{-1}$, corresponding to 3 × 3 × 3 k-point meshes for a cubic cell with the side length of 12.9 Å. For calculations with structure relation, the ionic positions, cell volume, and cell shape were treated as degrees of freedom (full relaxation). The convergence criteria for structure relaxation were set to $10^{-6}$ eV, and the force components were relaxed to $10^{-2}$ eV/Å. \\
The \textit{ab initio} molecular dynamics (AIMD)x calculations were done with the timestep of 1 fs, giving approximately 120 timesteps per thermalization from 0 to 2000 K. A friction parameter of 20 ps$^{-1}$ was used for each atom type, and the friction parameter of the lattice was set to 5 ps$^{-1}$. Each AIMD timestep was included in the MLIP training process. The crystal structures were imported from the Materials Project \cite{jain_commentary}, and fully relaxed using the DFT accuracy parameters used for MG calculations. Later, those relaxed structures are expanded/compressed or distorted, as described in the ``DFT database overview".
All the DFT calculations were done using the \textit{Intel Xeon Gold 6248} or \textit{Xeon Gold 6148} processors.
\\
For calculated MG systems with a number of atoms $N=150$, the average computational time of one single-point DFT calculation was 113 CPU hours, while the average DFT-minimization took 23 times longer (2621 CPU hours) and AIMD trajectory 97 times longer (11012 CPU hours). Therefore, even with a comparable number of structures in the MLIP train dataset to other approaches~\cite{PhysRevB.104.104101}, the developed methodology significantly shortens the MLIP development time.

\textcolor{darkgreen}{
\textit{Viscosity calculation ---} 
We perform NPT simulation for the 4500 particles and 20 independent runs at $P\approx 1$ bar and range of temperatures T$\in[1413,~1633]$. The viscosity is given by the Green-Kubo relation~\cite{chen2021breakdown}:
\begin{equation}
    \eta_{\alpha\beta} = \frac{V}{k_B T} \int_0^{\infty} P_{\alpha \beta}(t) P_{\alpha \beta}(0) dt 
\end{equation}
where $\{\alpha,\beta\} \in [xy, yz, zx]$, $T$ is the temperature, $k_B$ is the Boltzmann constant, and $V$ is the system volume.
The stress autocorrelation function is computed from well-relaxed NVT trajectories, which are run long enough to ensure convergence of the integral. Autocorrelation is calculated for the off-diagonal components and averaged to obtain the viscosity.}

\textit{Neuroevolutional Potential---} 
For training the MLIP, we use a neuroevolution potential NEP~\cite{fan2021neuroevolution,Fan_2022, song2024general} working on GPUs, within the GPUMD software. 
These potentials use a state-of-the-art evolutionary algorithm, the separable natural evolution strategy, to avoid local minima and yield robust parameter optimization~\cite{yao1999evolving}. 
For principles of the NEP model see Refs.~\cite{fan2021neuroevolution,fan2022improving}. The descriptor vectors used to describe the PEL include radial descriptors and angular descriptors.
During the training of the model, the loss function is minimized. It is defined as the weighted sum over the loss terms associated with energies, forces, and virials as well as the L1 and L2 norms of the parameter vector. For our trained MLIP such contributions are shown in Fig.~\ref{fig:dft_database_architecture_model_training}d) converging after around $10^5$ generations. A more detailed description of the loss function contributions can be found in the SI, Section~3.
The NEP4 version was used, with the default training parameters. The repulsive ZBL potential term \cite{Ziegler_the_stopping} was added to prevent particle overlap. The outer cutoff for the ZBL potential was set to 1.8~\AA, corresponding to the first coordination shell in the system. 
The radial, and angular cutoff were equal to 6.5~\AA, and 4~\AA, respectively. The former corresponds to the radial cutoff used in the EAM potential ~\cite{Cheng_2009}, while the latter is the default suggested in the GPUMD documentation \cite{GPUMD_documentation}. Both radial and angular descriptors were built with 8 basis functions, and the hidden layer consisted of 30 neurons. The training process was set to last $10^{5}$ generations (steps), which took about 14 hours on two \textit{Tesla V100-SXM2-32GB} Graphic Processing Units (GPUs). 
The NEP ecosystem does not need external dependence like Pytorch or TensorFlow. The trained MLIP can be directly extracted as a tabulated file and used directly in LAMMPS~\cite{LAMMPS} for MD simulations. 
All simulations comparing MLIP, and EAM were performed with LAMMPS, supported with the GPUMD~\cite{Fan_2022, song2024general} NEP interface. 
%\AW{The efficiency of those computations with MLIP and EAM is compared in SI Fig.~S4.}

%For the well relaxed samples, we perfrom the NVT simulations, suffuciently long

%First, 20 independent samples, each consisting of 4500 atoms arranged in a random structure, were generated. Each initial configuration underwent the following simulation protocol: NPT simulations were conducted at a pressure of $p$=1.01325~bar and at six different temperatures: 1633, 1593, 1553, 1513, 1473, 1413~K. Each simulation lasted 250000 time steps with a time step of $dt = 0.002$~ps. From the NPT simulations, the average system volumes were determined and used to rescale the simulation box sizes of the final configurations. Subsequently, NVT simulations were performed in two stages. The first stage aimed to equilibrate the system and bring it to thermal equilibrium (the same duration as the preceding NPT runs). In the second stage, lasting 2.2 ns, data of the pressure tensor were collected for calculating the viscosity given by the Green-Kubo relation~\cite{chen2021breakdown}:
%\begin{equation}
%    \eta = \frac{V}{k_B T} \int_0^{\infty} P_{\alpha \beta}(t) P_{\alpha \beta}(0) dt 
%\end{equation}
%where $T$ indicates the temperature, $k_B$ the Boltzmann constant and $V$ the volume of the system. The autocorrelation function was calculated for the three off-diagonal pressure tensor components ($P_{xy}$, $P_{xz}$, $P_{yz}$), and averaged.

\section*{Code Availability}
The DFT database developed to train and test the MLIP can be found at the following NOMAD Repository: \url{https://doi.org/10.17172/NOMAD/2025.03.20-2
}.
The code for the LJ-surrogate model is not publicly available but may be made available to qualified researchers at the reasonable request to the corresponding authors. 
\section*{Acknowledgements}  
This work has been supported by the European Union Horizon 2020 research and innovation program under grant agreement no.~857470 and from the European Regional Development Fund via the Foundation for Polish Science International Research Agenda PLUS program grant No.~MAB PLUS/2018/8. JSW work was supported by the National Science Centre, Poland, under research project no UMO-2019/35/D/ST5/03526. SB thanks the National Science Center in Poland for SONATA BIS number DEC-2023/50/E/ST3/00569 and FIRST TEAM FENG.02.02-IP.05-0177/23 project, funded by the Foundation for Polish Science.
A.W. and S.B. acknowledge the support from COST ACTION CA22154 (DAEMON) by COST (European Cooperation in Science and Technology). The authors acknowledge the computational 
resources provided by the Aalto University School of Science “Science-IT” project. M.A. thanks CSC (Finland) for support via the project 2010169. J.B. acknowledge funding from the Research council of Finland through the OCRAMLIP project, grant number 354234.

\vspace{0.5cm}

\section*{Author Contributions}
A.D.S.P., S.B., J.B. and J.W. developed the methodology. A.W., A.D.S.P. and M.A. carried out the simulations and performed the data analysis. F.K. carried out MD simulations. S.B. conceptualized the research. All authors contributed to the writing of the manuscript.

\vspace{0.5cm}

\section*{Competing Interests}
%\vspace{-0.5cm}
The authors have no competing interests to declare.

\bibliography{biblio}

\end{document}

\title{Efficient training of machine learning potentials for metallic glasses: CuZrAl validation \\
Supplementary Information}

\author{Antoni Wadowski}
\affiliation{NOMATEN Centre of Excellence, National Center for Nuclear Research, ul. A. So\l{}tana 7, 05-400 Swierk/Otwock, Poland.}
\affiliation{Faculty of Materials Science and Engineering, Warsaw University of Technology, Wo\l{}oska 141, 02-507 Warsaw, Poland.}

\author{Anshul D.S. Parmar}
\email{Anshul.Parmar@ncbj.gov.pl}
\affiliation{NOMATEN Centre of Excellence, National Center for Nuclear Research, ul. A. So\l{}tana 7, 05-400 Swierk/Otwock, Poland.}

\author{Filip Kaśkosz}
\affiliation{NOMATEN Centre of Excellence, National Center for Nuclear Research, ul. A. So\l{}tana 7, 05-400 Swierk/Otwock, Poland.}

\author{Jesper Byggm\"astar}
\affiliation{Department of Physics, P.O. Box 43, FI-00014 University of Helsinki, Finland.}

\author{Jan S. Wr\'obel}
\affiliation{Faculty of Materials Science and Engineering, Warsaw University of Technology, Wo\l{}oska 141, 02-507 Warsaw, Poland.}

\author{Mikko J. Alava}
\affiliation{NOMATEN Centre of Excellence, National Center for Nuclear Research, ul. A. So\l{}tana 7, 05-400 Swierk/Otwock, Poland.}
\affiliation{Aalto University, Department of Applied Physics, PO Box 11000, 00076 Aalto, Espoo, Finland.}

\author{Silvia Bonfanti}
\email{Silvia.Bonfanti@unimi.it}
\affiliation{NOMATEN Centre of Excellence, National Center for Nuclear Research, ul. A. So\l{}tana 7, 05-400 Swierk/Otwock, Poland.}
\affiliation{Center for Complexity and Biosystems, Department of Physics "Aldo Pontremoli", University of Milan, Via Celoria 16, 20133 Milano, Italy.}

\date{\today}

\maketitle
Here we provide additional information about the following topics: 1.~DFT-relaxation effects,
2.~MLIP vs EAM performance, and 
3.~MLIP training details.

\section{1. DFT-relaxation effects}
\begin{figure}[htbp]
    \centering
    \includegraphics[width=0.5\textwidth]{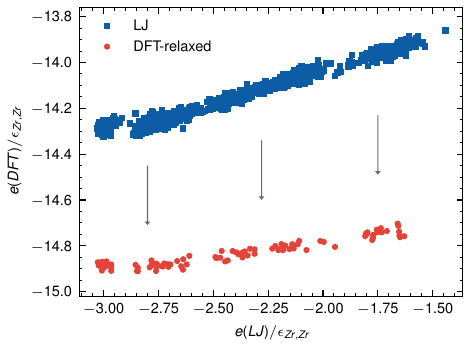}
    \caption{
    DFT energy e(DFT) of the structures within the train and test DFT datasets, as a function of the LJ energy e(LJ) of the LJ-surrogate-model samples. The subset of those samples (blue squares) is subjected to DFT minimization (red circles), and the resulting e(DFT) values change is shown. The DFT relaxation process is symbolized by the vertical arrows.
    }
    \label{fig:dft_relaxation_energy_effects} 
\end{figure}

{\bf Exploring energy landscape---} Figure ~\ref{fig:dft_relaxation_energy_effects} underlines the linear relationship between the LJ-surrogate PEL and the underline landscape defined by the DFT calculation. The relationship holds the ``instantaneous-'' liquid-like structure and the DFT relaxed energy minimum structures, too. Such observations confirm the methodology's effectiveness reported in the main manuscript.

\section{2. MLIP vs EAM performance}

\begin{figure}[h]
    \centering
    \includegraphics[width=0.46\textwidth]{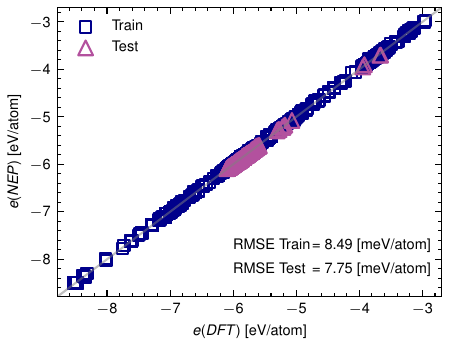}
    \includegraphics[width=0.46\textwidth]{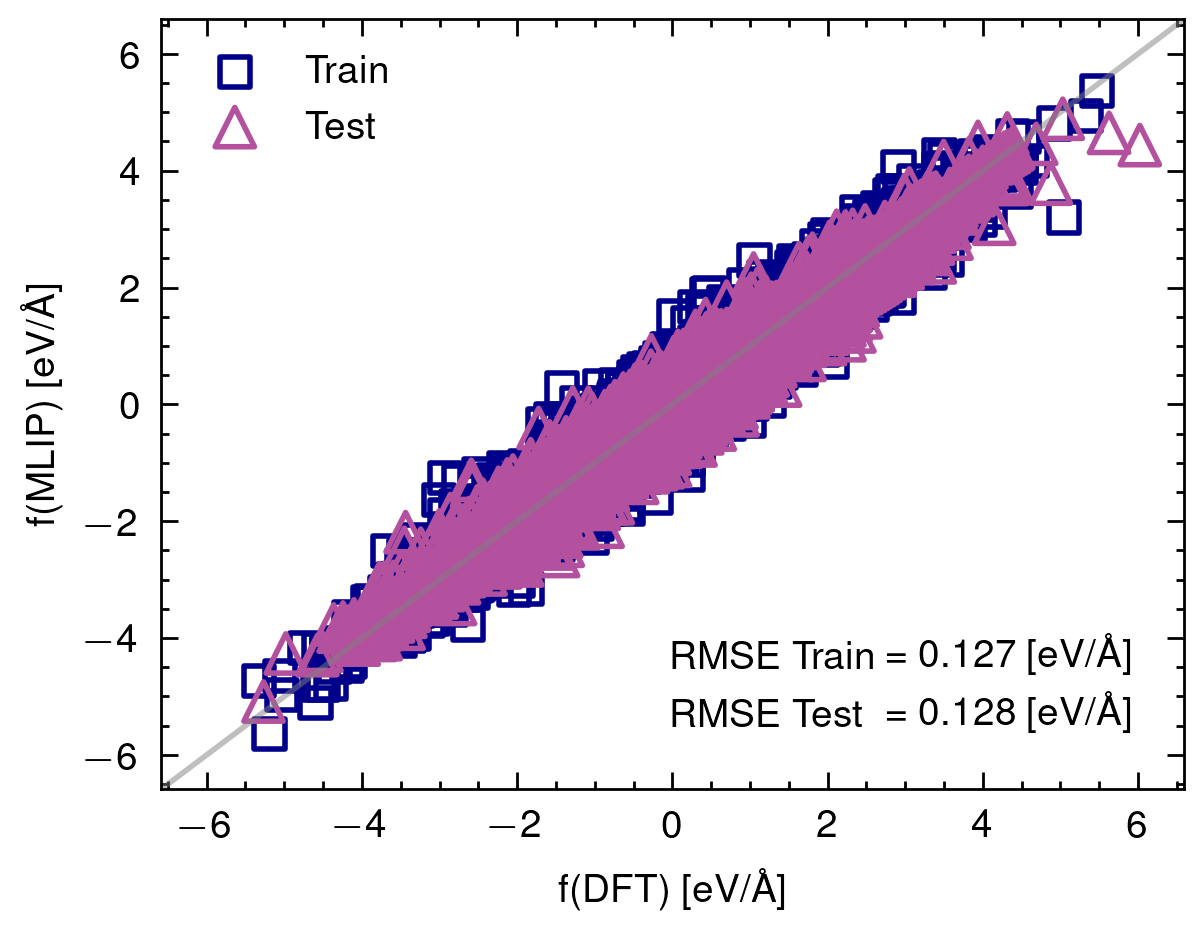}
    \includegraphics[width=0.46\textwidth]{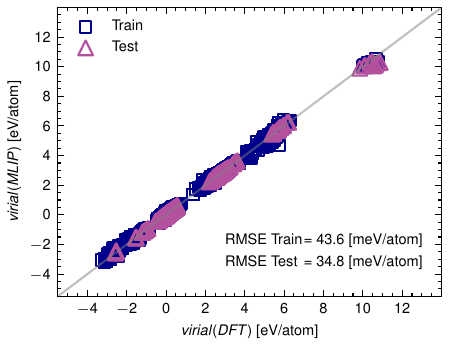}
    \includegraphics[width=0.46\textwidth]{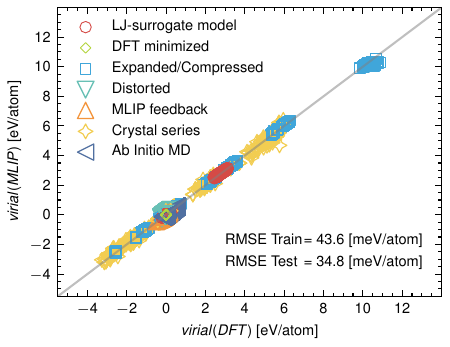}
    \caption{Comparison of the MLIP prediction vs DFT data. For both training (dark blue square) and test (purple triangle) datasets, the MLIP's output is shown for~\textbf{a)} energies, \textbf{b)} force vector $x$, $y$, $z$ components, and~\textbf{c)} virial matrix $[xx,yy,zz,xy,yz,zx]$ components. The same virial data is shown in~\textbf{d)}, and coded by structure type.}
    \label{fig:training_loss}
    \vspace{-0.5cm}
\end{figure}

{\bf Accuracy---}
Figures~\ref{fig:training_loss}a),b),c) report the MLIP's performance for the test (triangles), and train (squares) dataset, indicating satisfactory generalization of the developed MLIP. Moreover, the the virial predictions are similarly accurate for all subsets of the DFT database, as shown in Fig.~\ref{fig:training_loss}d). This observation is consistent with the corresponding energy and force data (Fig.~3b) and 3c)). This denotes the model's precision in predicting diverse material attributes while demonstrating an absence of bias toward specific configurations.

\section{3. MLIP training details}

{\bf Neuroevolution potential (training algorithm procedure)---}
The loss function minimized during neuroevolution potential training has this form~\cite{Fan_2022}.
\begin{equation}
    \begin{split}L(\boldsymbol{z})
    &= \lambda_\mathrm{e} \left(
    \frac{1}{N_\mathrm{str}}\sum_{n=1}^{N_\mathrm{str}} \left( U^\mathrm{NEP}(n,\boldsymbol{z}) - U^\mathrm{tar}(n)\right)^2
    \right)^{1/2}  \\
    &+  \lambda_\mathrm{f} \left(
    \frac{1}{3N}
    \sum_{i=1}^{N} \left( \boldsymbol{F}_i^\mathrm{NEP}(\boldsymbol{z}) - \boldsymbol{F}_i^\mathrm{tar}\right)^2
    \right)^{1/2}  \\
    &+  \lambda_\mathrm{v} \left(
    \frac{1}{6N_\mathrm{str}}
    \sum_{n=1}^{N_\mathrm{str}} \sum_{\mu\nu} \left( W_{\mu\nu}^\mathrm{NEP}(n,\boldsymbol{z}) - W_{\mu\nu}^\mathrm{tar}(n)\right)^2
    \right)^{1/2}  \\
    &+  \lambda_1 \frac{1}{N_\mathrm{par}} \sum_{n=1}^{N_\mathrm{par}} |z_n|  \\
    &+  \lambda_2 \left(\frac{1}{N_\mathrm{par}} \sum_{n=1}^{N_\mathrm{par}} z_n^2\right)^{1/2}
    \end{split}
\label{eq:loss_function}
\end{equation}

\begin{textblock*}{1cm}(2cm,1.8cm)
    \textbf{a)}
\end{textblock*}
\begin{textblock*}{1cm}(10.8cm,1.8cm)
    \textbf{b)}
\end{textblock*}
\begin{textblock*}{1cm}(2cm,8.3cm)
    \textbf{c)}
\end{textblock*}
\begin{textblock*}{1cm}(10.8cm,8.3cm)
    \textbf{d)}
\end{textblock*}
where $N_{str}$, and is a number of structures in the training dataset for the full batch or the number of structures in the mini-batch. $N$ is the number of atoms in each structure, $i$ is $i$th atom in the structure, and $z$ denotes the neural network parameters. The first three terms represent root mean square errors (RMSEs) between the NEP predictions (NEP) in the current training generation and the target values (tar). The RMSEs are calculated for the energies $U$, forces $F$, and virials $W_{\mu\nu}$. The last two terms correspond to the $\mathcal{L}_1$, and $\mathcal{L}_2$ regularization terms of the parameter vector. The weights $\lambda_\mathrm{e}$ are a tunable hyper-parameters.

\bibliography{biblio}